\documentclass[useAMS,usenatbib,usegraphicx]{mn2e}

\title[On the magnetic fields in voids]{On the magnetic fields in voids}
\author[A. M. Beck et al.]{A. M. Beck$^{1,2}$\thanks{E-mail: abeck@usm.uni-muenchen.de}, M. Hanasz$^{3}$, H. Lesch$^{1}$, R.-S. Remus$^{1,2}$ and F. A. Stasyszyn$^{1}$\\
  $^{1}$University Observatory Munich, Scheinerstr. 1, D-81679 Munich, Germany\\
  $^{2}$Max Planck Institute for Extraterrestrial Physics, Giessenbachstr., D-85748 Garching, Germany\\
  $^{3}$Toru\'{n} Centre for Astronomy, Nicolaus Copernicus University, PL-87-148 Piwnice/Toru\'{n}, Poland}

\usepackage[latin1]{inputenc}
\usepackage{units}
\usepackage{float}
\usepackage{graphicx}
\usepackage{amsmath}
\usepackage{amssymb}
\usepackage{amsfonts}
\usepackage{xcolor}
\usepackage{times}
\usepackage{comment}

\begin{document}

\date{Accepted XXXX. Received XXXX; in original form XXXX}

\pagerange{\pageref{firstpage}--\pageref{lastpage}} \pubyear{2012}

\maketitle

\label{firstpage}


\begin{abstract}
\noindent{}We study the possible magnetization of cosmic voids by void galaxies.
Recently, observations revealed isolated starforming galaxies within the voids.
Furthermore, a major fraction of a voids volume is expected to be filled with magnetic fields of a minimum strength of about $10^{-15}$ G on Mpc scales.
We estimate the transport of magnetic energy by cosmic rays (CR) from the void galaxies into the voids.
We assume that CRs and winds are able to leave small isolated void galaxies shortly after they assembled, and then propagate within the voids.
For a typical void, we estimate the magnetic field strength and volume filling factor depending on its void galaxy population and possible contributions of strong active galactic nuclei (AGN) which border the voids.
We argue that the lower limit on the void magnetic field can be recovered, if a small fraction of the magnetic energy contained in the void galaxies or void bordering AGNs is distributed within the voids.
\end{abstract}




\section{Introduction}
\noindent{}Recently, high energy observations revealed a lower limit of about $10^{-15}$ G on Mpc lengths for cosmic scale magnetic fields.
These observations indicate the existence of magnetic fields in voids, with an argumentation as follows.
TeV $\gamma$-ray photons from distant extragalactic blazars are passing through voids, creating electron/positron pairs when interacting with the extragalactic background light.
These pairs would travel in the same direction as the original photon and produce an observable electromagnetic cascade emission.
However, in the presence of void magnetic fields, the pairs are deflected and the cascade emission is suppressed.
Since the observations of the distant TeV blazars do not detect the full cascade emission, magnetic fields have to be present in at least half of a voids volume \citep[see e.g.][]{neronov10,tavecchio10,dermer11,dolag11,essey11,huan11,tavecchio11,taylor11,arlen12,essey12,kusenko12,miniati12b,neronov12,takahashi12}.

\noindent{}Of course, the origin of magnetic fields in the empty voids appears to be enigmatic.
We do not intend to discuss the several proposals, but rather point out the new perspectives given by the most recent detections of a galaxy population in the voids themselves.
Until recently, voids have been considered as completely empty regions, present in the largest structures known in our Universe in a web-like distribution.
This cosmic web is the result of anisotropies during the gravitational collapse within an expanding Universe, when matter gets concentrated within overdense regions, the filaments and sheets \citep[see e.g.][and references therein]{mo10}.

\noindent{}Recently, first public void catalogues of the local Universe have been constructed \citep{pan12,sutter12}.
These surveys, performed on the Sloan Digital Sky Survey Data Release 7 \citep{abazajian09}, show that voids tend to have elliptical shapes and a high density contrast at the borders.
Their effective radii range from a few Mpc up to several hundred Mpc.

\noindent{}However, voids are only less dense regions and still contain matter and structures.
Over the past years, several void galaxy surveys have been performed identifying and analysing galaxies within the voids \citep[see e.g.][]{grogin99,grogin00,rojas04,rojas05,park07,kreckel11a,kreckel11b,pustilnik11,hoyle12,tavasoli12}.
From the SDSS DR 7, \cite{pan12} identified a sample of $\approx{}10^{3}$ voids, hosting $\approx{}10^{4}-10^{5}$ galaxies.

\noindent{}Surprisingly, the void galaxies are similar to the corresponding galaxies in the high density environments \citep[see e.g.][]{kreckel12}.
They tend to be blue galaxies and exhibit effective radii of a few kpc, but are less massive and lower in luminosity.
Also, they are commonly gasrich, starforming and show a regular rotation, however, most have disturbed gas morphologies indicating ongoing accretion or strong turbulence.
The galaxies live mainly in isolation and evolve slowly, but a few appear in small groups.

\noindent{}We may summarize that a typical void contains a few ten starforming galaxies.
Now, we can design a scenario, which relies on the following line of thoughts derived from the evolution of magnetic fields in well-studied galaxies \citep[see also][]{kronberg99,samui08,chyzy11}.

\noindent{}Galaxies in the process of assembly are known to build up an equipartition magnetic field \citep[on cosmic magnetism see e.g.][]{kulsrud08,vallee11}.
It is assumed that first supernova explosions deliver interstellar magnetic seed fields of the order of $10^{-9}$ G \citep[see e.g.][]{bisnovatyi73,rees06}.
During the galactic halo and galaxy assembly the magnetic field is amplified up to equipartition with the corresponding turbulent energy density by small-scale dynamo action \citep[see e.g.][]{kulsrud97,beck12,geng12a,geng12b}.
Turbulence yields a total amplification time of a few hundred Myrs, leading to a $\mu$G magnetic field very shortly after the assembly process started.
Even if the galaxies evolve slowly, the existence of equipartition magnetic fields can be assumed at high redshift \citep[see e.g.][]{zweibel06,kronberg08,arshakian09,beck12}.

\noindent{}The star formation within the forming void galaxies also leads to the production of CRs within supernovae and acceleration of CRs within supernova remnants \citep[see e.g.][]{longair10}.
A dynamo driven by these CRs contributes to the amplification of the galactic-scale magnetic field \citep{lesch03,hanasz09,siejkowski10}.
Furthermore, CRs are driving winds from the galaxies.
CR-driven winds can attain high velocities exceeding the escape velocity of galactic haloes and therefore the CRs could propagate into the voids \citep[see e.g.][]{bertone06,breitschwerdt08,everett08,samui10,ensslin11,uhlig12,dorfi12}.
Together with the CRs escaping from the galaxies, magnetic field lines are carried outwards, resulting in the transport of magnetic energy into the voids \citep[see e.g.][]{longair10}.
We note that the electric current carried by the propagating CRs, may generate magnetic fields at a rate of $10^{-16}$ G/Gyr within the voids \citep{miniati11,miniati12a}.
However, for significantly stronger magnetic fields, different mechanisms or the transport of magnetic energy together with the CRs are necessary.

\noindent{}Furthermore, black holes are commonly assumed to reside at the center of galaxies.
These black holes are known to launch jets of charged particles, which can transport magnetic fields far into the intergalactic medium, in the case of supermassive black holes within giant radio galaxies even onto scales of several Mpc \citep[see e.g.][]{willis78,strom80,kronberg94,kronberg01,kronberg09,colgate11}.
However, the small mass of the void galaxies makes a void supermassive black hole population unlikely.
Also, \cite{kreckel12} did not find evidence for strong AGN activity within their sample of void galaxies.
Hence, a magnetisation of the voids by an intrinsic population of supermassive black holes seems not plausible.
Still, dwarf galaxies can host intermediate mass black holes \citep[see e.g.][]{bellovary11,nyland12}, whose pc or kpc scale jets support the outflows and winds.
In addition, the highly magnetized Mpc scale jets of strong AGNs at the voids borders can penetrate into the voids and contribute to their magnetisation.

\noindent{}In this letter, we combine the latest observations of void magnetic fields and void galaxies.
We discuss the transport of magnetic energy from the void galaxies and bordering AGNs into the voids by CRs.


\section{Estimations}

\noindent{}Before starting with the estimations, we will obtain some characteristic values.
From a public void catalogue \citep{pan12,sutter12}, we find that typical voids have a characteristic radius of $R_\rmn{V}\approx{}20$ Mpc and contain $N\approx{}10$ starforming galaxies.
Within the voids, magnetic fields of at least $B_\rmn{V}\approx{}10^{-15}$ G have been detected \citep[see e.g.][]{neronov10}.
From the void galaxy survey \citep[see e.g.][]{kreckel12}, we find a typical void galaxy to have a characteristic radius of $R_\rmn{G}\approx{}3$ kpc and, if star formation was constant, an age of $T_\rmn{G}\approx{}7.5$ Gyrs.
The dynamical mass is lower than $M_\rmn{G}\approx{}10^{11}$ $M_{\odot}$ and the velocity dispersion $\sigma\approx{}150$ km/s, leading to a galactic halo virial radius of $R_\rmn{H}=GM/\sigma^2\approx{}60$ kpc.
We assume the galactic equipartition magnetic field to be $B_\rmn{G}\approx{}5$ $\mu$G \citep[see e.g.][]{vallee11}.

\noindent{}First, we can compare the magnetic energies of a typical void and of a typical void galaxy.
The magnetic energy within a sphere of radius R and with a magnetic field $B$ is given by
\begin{equation}E=\frac{R^3B^2}{6}.\end{equation}
Thus, the magnetic energy within the galaxy is of the order $E_\rmn{G}\approx{}10^2(\mu\rmn{G})^2\rmn{kpc}^3$ and of the void at least of the order $E_\rmn{V}\approx{}10^{-6}(\mu\rmn{G})^2\rmn{kpc}^3$.
The void contains only a fraction of the magnetic energy produced within a void galaxy.
This fraction can also be recovered, when expanding the magnetic energy from the galaxy radius onto the void radius (i.e. $(R^3_\rmn{G}/R^3_\rmn{V})^{2/3}\approx{}10^{-8}$).
Therefore, the magnetic energy contained within a void galaxy is sufficient in magnetizing the void at the observed level, if it can be transported far enough outwards.
We will argue that CRs are responsible for the magnetisation of a fraction of the voids volumes.
The propagation of CRs carrying an electric current through space can already generate magnetic fields at a rate of $10^{-16}$ G/Gyr \citep{miniati11}.
However, in addition to that, we note that magnetic field lines from the void galaxies can be dragged along with the propagating CRs into the voids, leading to the transport of magnetic energy.

\noindent{}The CRs are mainly produced in supernovae and accelerated within supernova remnants.
Intermediate mass black holes at the centers of the void galaxies can drive supersonic outflows, leading to shocks that subsequently accelerate charged particles.
The confinement time of CRs within galaxies is known to be of the order of a few ten Myrs \citep[see e.g.][]{longair10}.
The escape velocity of a galactic halo depends on its mass, and the distance from the center of mass, at which the particles are launched.
For galaxies in the process of assembly with yet small masses, the winds are launched further outside, as the mass is not yet compressed within a central region.
Numerical simulations indicate that cosmic-ray driven winds exceed the escape speed of dwarf galactic haloes \citep{samui10,uhlig12}.
These numerical models also show, that the smaller the haloes, the more spherically symmetric the outflows.
Recent work by \cite{dorfi12} finds that time-dependent effects of winds and shocks within the galactic haloes could reaccelerate the CRs, leading to wind speeds exceeding the escape velocity.
It is known that CRs can be confined within group atmospheres, if the group is large enough and the energy of the CRs is too small \citep[see][for details]{berezinsky97}.
However, for the atmospheres of small dwarf galaxy groups, particles with an energy of 1 GeV can still escape.
We note that if the void galaxies are grouped into too large structures, the CRs and hence also the winds can be confined and not escape into the voids.
Summing up, if void galaxies are smaller in mass and reside mainly in isolation, a spherically symmetric galactic wind can escape into the voids and propagate within.

\noindent{}Now, we ask with which velocity the CRs are propagating within the voids.
We estimate this velocity by considering the lowest possible diffusion coefficient (i.e. Bohm diffusion) for particle propagation along magnetic field lines
\begin{equation}D_\rmn{Bohm}=\frac{c}{3}r_\rmn{Gyro}=\frac{c}{3}\frac{\gamma{}m_\rmn{p}c^{2}}{eB},\end{equation}
with the speed of light $c$, the elementary charge $e$, the proton mass $m_\rmn{p}$ and a Lorentz factor $\gamma$.
For energies as low as 1 GeV and a length scale of the galactic halo size, the diffusion speed $V_\rmn{CR}=D_\rmn{Bohm}/R_\rmn{H}$ reaches values of $V_\rmn{CR}\approx{}1500$ km/s at the galactic peripheries.
Because the Bohm diffusion coefficient represents the slowest possible diffusion of charged particles along magnetic field lines, the propagation velocity of the CRs can be higher.
The CRs can propagate far by themselves forming bubbles around the void galaxies, whose expansion, over time, will also be supported by the Hubble flow.
However, for simplicity, we assume that CRs are propagating at a speed of at least 1500 km/s or 1.5 Mpc/Gyr \citep[see also][]{miniati11}.

\noindent{}Next, we want to estimate how far CRs could have propagated since the assembly of the void galaxies started.
If, roughly, the build-up of magnetic fields by a turbulent dynamo and the propagation of CRs to the galactic periphery took one Gyr, CRs could have still been propagating within the voids for $\approx{}6.5$ Gyrs.
With our assumption of the propagation velocity, this would lead to a travelled distance of the CRs of $R_\rmn{B}\approx{}10$ Mpc.
For simplicity, we assume the expansion to be spherically symmetric.
This then allows us to estimate the volume filling factor of a typical void, which, when assuming a population of N randomly distributed galaxies within, is
\begin{equation}f=\sqrt{N}\left(\frac{R_\rmn{Bubble}}{R_\rmn{Void}}\right)^{3}.\end{equation}
At the start of the galactic assembly process, the voids have a negligible volume filling factor.
However, if a typical present-day void hosts ten randomly distributed galaxies, the voids volume filling factor would be about $\approx{}0.4$.
We note that this estimation is highly speculative, as the alignment of the galaxies within the voids as well as the propagation speed and hence the propagated distances are subject to large uncertanties.

\noindent{}Last, we want to approximate the magnetic field strength within a typical void.
By expanding the galactic magnetic field into the bubbles driven by the propagating CRs, we estimate
\begin{equation}B_\rmn{Bubble}=\epsilon{}\sqrt{N}B_\rmn{Galaxy}\left(\frac{V_\rmn{Galaxy}}{V_\rmn{Bubble}}\right)^{\frac{2}{3}},\end{equation}
where $\epsilon$ is the fraction of magnetic energy transported from the void galaxies into the voids and N the number of randomly distributed galaxies within a typical void.
With our characteristic values, and choosing a very low fraction of $\epsilon=0.001$, we recover the observed lower limit on the void magnetic field of $10^{-15}$ G.
However, numerical simulations performed by \cite{siejkowski10} indicate the outflow fraction of magnetic energy to be much higher, which would lead to void magnetic fields significantly stronger than the detected limit.

\noindent{}We note that the above estimations can also be used to approximate the contributions of AGN lobes on the void magnetic fields, if $N$ is assumed to be the number of randomly placed lobes inside a void, and $B$ and $V$ the characteristic lobe magnetic field strength and volume,
From \cite{kronberg01} we take for the lobe of a typical strong AGN a characteristic field strength of the order of $B_\rmn{AGN}\approx{}5\mu$G within a volume of about $V_\rmn{AGN}\approx{}(250 \rmn{kpc})^{3}$.
These values give a magnetic energy of about $E_\rmn{AGN}\approx{}10^7(\mu\rmn{G})^2\rmn{kpc}^3$ inside a lobe, a value which is many orders of magnitude higher than the lower limit magnetic energy contained within a typical void.
We can then approximate the contribution of just one strong AGN lobe, which we assume to have been placed into the void ten Gyrs ago.
With the above Bohm diffusion speed the lobe would contribute to the present-day magnetic filling factor by $\approx{}0.4$.
Choosing a very small fraction of $\epsilon=10^{-5}$, we again recover the observed lower limit on the void magnetic field of $10^{-15}$ G.
Furthermore, distributing the entire magnetic energy of a strong AGN lobe within a void would also lead to void magnetic fields significantly stronger than the detected limit.


\section{Summary}
\begin{figure}
\begin{center}
  \includegraphics[width=0.475\textwidth, bb = 254 336 620 973]{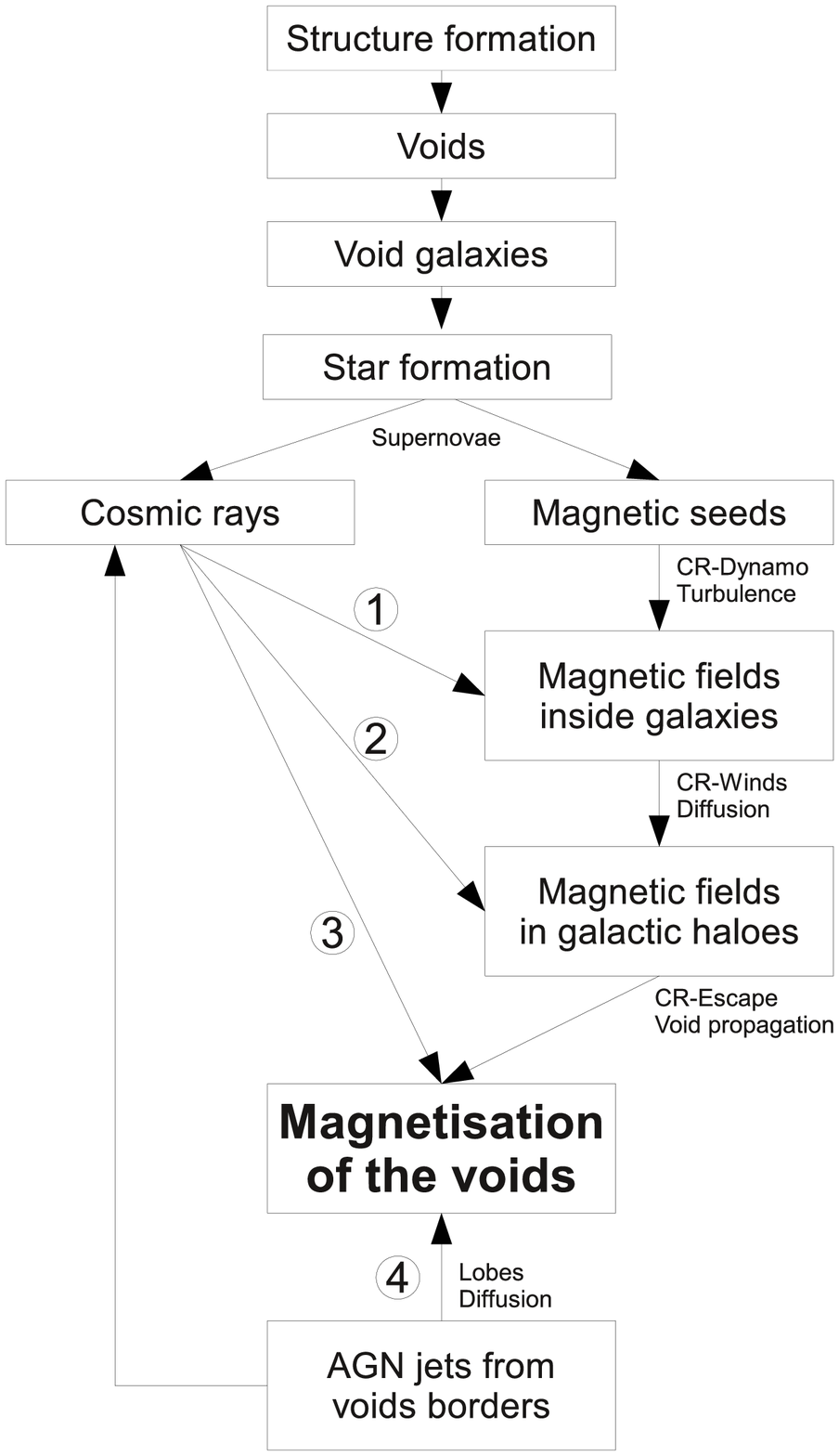}
  \caption{Schematic view on the processes and structures involved in the magnetisation of cosmic voids.
CRs lead to the amplification of the void galactic magnetic field (1), as well as winds escaping into the void galactic haloes (2) and the magnetization of the voids themselves (3), supported by AGNs which border the voids (4).}
  \label{fig:structure}
\end{center}
\end{figure}

\begin{figure*}
\begin{center}
  \includegraphics[width=0.95\textwidth, bb = 10 40 710 500]{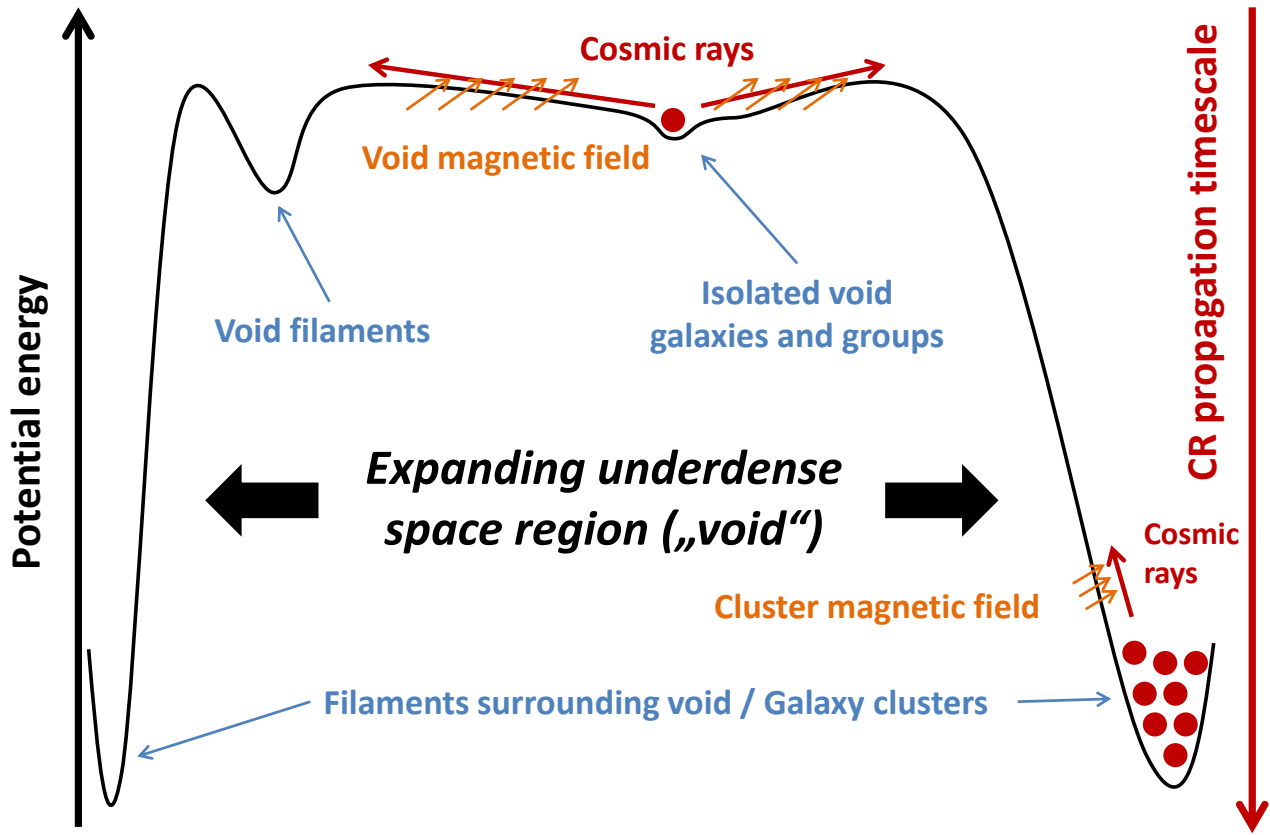}
  \caption{Schematic view on a typical void and its galaxies within, as well as the connected CRs and magnetic fields.
CRs are able to propagate more easy into the voids from the void galaxies than from the filament galaxies.}
  \label{fig:schematic}
\end{center}
\end{figure*}

\noindent{}Figures \ref{fig:structure} and \ref{fig:schematic} illustrate our argumentation, summarized as follows.
Voids are underdense space regions, growing over cosmic time and surrounded by large filaments.
However, within the voids are void galaxies, dwarf-like, mainly isolated and evolving slowly.
The galaxies appear to be similar to the corresponding galaxies from high density environments.
They undergo star formation, leading to supernovae and the production of CRs and magnetic seed fields.
Fast turbulent and CR-dynamos are able to build up an equipartition magnetic field during the early stages of the galaxy assembly.
The CRs are driving high-velocity winds escaping from the galaxies into the voids.
The small mass of the void galaxies makes strong AGN activity inside the voids unlikely, but intermediate mass black holes can support the winds.
The CRs and the winds are able to leave the atmospheres of the void galactic haloes, in contrast to galactic haloes residing within large filaments or clusters.
Over time, the CRs can propagate far into the voids, magnetising a fraction of the voids volumes.
The voids are growing by cosmic expansion, but the propagation of the CRs also increases the magnetised volume fraction.
When assuming expansion of a fraction of the void galactic magnetic fields into bubbles driven by CRs, a lower limit for the magnetic field of $>10^{-15}$ G can be recovered, and even stronger magnetic fields are possible.

\noindent{}We have presented and discussed a qualitative scenario for the origin of the magnetic fields in voids.
There are two major contributors to the void magnetic fields.
First, an intrinsic galaxy population within the voids can produce and spill out magnetic fields.
Second, highly magnetized jets from giant radio galaxies bordering the voids can penetrate the voids.
Both mechanisms are capable of delivering enough magnetic energy into the voids to yield the observed lower limit of void magnetic fields.
However, the high volume filling factors especially for the largest voids still remain challenging.
In the view without primordial magnetic seed fields, a combined contribution of void galaxies and border AGNs is probably responsible for the magnetic fields in voids.

\section*{Acknowledgments}
\noindent{}AMB is deeply grateful for the hospitality of the University of Konstanz, and many discussions with Annette Geng, Marcus Beck and Peter Nielaba.
We thank our referee P.~P. Kronberg for prompt reviewing and valuable suggestions.
FAS is supported by the DFG Research Unit FOR1254.
MH acknowledges the generous support of the Alexander von Humboldt foundation during his stay at the University Observatory Munich.
RSR acknowledges a grant from the International Max-Planck Research School of Astrophysics (IMPRS).


\bsp

\label{lastpage}

\end{document}